\documentclass[master]{ws-rv9x6} 
\usepackage{ws-rv-har}    
\usepackage{ws-rv-thm}    
\usepackage{subfigure}    
\usepackage{ws-index}     
\usepackage{enumitem}
\usepackage[percent]{overpic}
\usepackage{xcolor}
\makeindex
\newindex{aindx}{adx}{and}{Author Index}       
\renewindex{default}{idx}{ind}{Subject Index}  

\begin{document}		
\newcommand{\ltsima}{$\; \buildrel < \over \sim \;$}
\newcommand{\lsim}{\lower.5ex\hbox{\ltsima}}
\newcommand{\gtsima}{$\; \buildrel > \over \sim \;$}
\newcommand{\gsim}{\lower.5ex\hbox{\gtsima}}
\newcommand{\bra}{\langle}
\newcommand{\ket}{\rangle}
\newcommand{\lprime}{\ell^\prime}
\newcommand{\lpp}{\ell^{\prime\prime}}
\newcommand{\mprime}{m^\prime}
\newcommand{\mpp}{m^{\prime\prime}}
\newcommand{\ci}{\mathrm{i}}
\newcommand{\dd}{\mathrm{d}}
\newcommand{\veck}{\mathbf{k}}
\newcommand{\vecx}{\mathbf{x}}
\newcommand{\vecr}{\mathbf{r}}
\newcommand{\vecv}{\mathbf{\upsilon}}
\newcommand{\vecw}{\mathbf{\omega}}
\newcommand{\vecj}{\mathbf{j}}
\newcommand{\vecq}{\mathbf{q}}
\newcommand{\vecl}{\mathbf{l}}
\newcommand{\vecn}{\mathbf{n}}
\newcommand{\lm}{\ell m}
\newcommand{\that}{\hat{\theta}}
\newcommand{\thatp}{\that^\prime}
\newcommand{\chip}{\chi^\prime}
\newcommand{\hs}{\hspace{1mm}}
\newcommand{\nar}{New Astronomy Reviews}
\def\gsim{~\rlap{$>$}{\lower 1.0ex\hbox{$\sim$}}}
\def\lsim{~\rlap{$<$}{\lower 1.0ex\hbox{$\sim$}}}
\def\Msun {\,\mathrm{M}_\odot}
\def\Jcrit {J_\mathrm{crit}}
\newcommand{\rsun}{R_{\odot}}
\newcommand{\mbh}{M_{\rm BH}}
\newcommand{\Msunyr}{M_\odot~{\rm yr}^{-1}}
\newcommand{\mdot}{\dot{M}_*}
\newcommand{\ledd}{L_{\rm Edd}}
\newcommand{\cmc}{{\rm cm}^{-3}}
\def\gsim{~\rlap{$>$}{\lower 1.0ex\hbox{$\sim$}}}
\def\lsim{~\rlap{$<$}{\lower 1.0ex\hbox{$\sim$}}}
\def\Msun {\,\mathrm{M}_\odot}
\def\Jcrit {J_\mathrm{crit}}

\def\simgreat{\lower2pt\hbox{$\buildrel {\scriptstyle >}
   \over {\scriptstyle\sim}$}}
\def\simless{\lower2pt\hbox{$\buildrel {\scriptstyle <}
   \over {\scriptstyle\sim}$}}
\def\msobh{M_\bullet^{\rm sBH}}
\def\zodot{\,{\rm Z}_\odot}
\newcommand{\lambdabar}{\mbox{\makebox[-0.5ex][l]{$\lambda$} \raisebox{0.7ex}[0pt][0pt]{--}}}

\def\na{NewA}%
\def\aj{AJ}%
\def\araa{ARA\&A}%
\def\apj{ApJ}%
\def\apjl{ApJ}%
\def\jcap{JCAP}

\def\pasa{PASA}

\def\apjs{ApJS}%
\def\ao{Appl.~Opt.}%
\def\apss{Ap\&SS}%
\def\aap{A\&A}%
\def\aapr{A\&A~Rev.}%
\def\aaps{A\&AS}%
\def\azh{AZh}%
\def\baas{BAAS}%
\def\jrasc{JRASC}%
\def\memras{MmRAS}%
\def\mnras{MNRAS}%
\def\pra{Phys.~Rev.~A}%
\def\prb{Phys.~Rev.~B}%
\def\prc{Phys.~Rev.~C}%
\def\prd{Phys.~Rev.~D}%
\def\pre{Phys.~Rev.~E}%
\def\prl{Phys.~Rev.~Lett.}%
\def\pasp{PASP}%
\def\pasj{PASJ}%
\def\qjras{QJRAS}%
\def\skytel{S\&T}%
\def\solphys{Sol.~Phys.}%

\def\sovast{Soviet~Ast.}%
\def\ssr{Space~Sci.~Rev.}%
\def\zap{ZAp}%
\def\nat{Nature}%
\def\iaucirc{IAU~Circ.}%
\def\aplett{Astrophys.~Lett.}%
\def\apspr{Astrophys.~Space~Phys.~Res.}%
\def\bain{Bull.~Astron.~Inst.~Netherlands}%
\def\fcp{Fund.~Cosmic~Phys.}%
\def\gca{Geochim.~Cosmochim.~Acta}%
\def\grl{Geophys.~Res.~Lett.}%
\def\jcp{J.~Chem.~Phys.}%
\def\jgr{J.~Geophys.~Res.}%
\def\jqsrt{J.~Quant.~Spec.~Radiat.~Transf.}%
\def\memsai{Mem.~Soc.~Astron.~Italiana}%
\def\nphysa{Nucl.~Phys.~A}%

\def\physrep{Phys.~Rep.}%
\def\physscr{Phys.~Scr}%
\def\planss{Planet.~Space~Sci.}%
\def\procspie{Proc.~SPIE}%

\newcommand{\rmp}{Rev. Mod. Phys.}
\newcommand{\ijmpd}{Int. J. Mod. Phys. D}
\newcommand{\sovjetp}{Soviet J. Exp. Theor. Phys.}
\newcommand{\jkas}{J. Korean. Ast. Soc.}
\newcommand{\PPVI}{Protostars and Planets VI}
\newcommand{\njp}{New J. Phys.}
\newcommand{\rap}{Res. Astro. Astrophys.}



\setcounter{page}{99}
\setcounter{chapter}{4}


%
%
%
%
%
%
%
%
%
%
%
%
%
%
%


\title{Formation of the First Black Holes}
\chapter[Black hole formation via gas-dynamical processes]{Black hole formation via gas-dynamical processes$^1$ }\label{dcbh}
\author[Muhammad A. Latif ]{Muhammad A. Latif}
\address{Department of Physics, United Arab Emirates University,\\
Po Box, 15551, Al-Ain, UAE,\\ latifne@gmail.com}
\footnotetext{$^1$ Preprint~of~a~review volume chapter to be published in Latif, M., \& Schleicher, D.R.G., ''Black hole formation via gas-dynamical processes'', Formation of the First Black Holes, 2018 \textcopyright Copyright World Scientific Publishing Company, https://www.worldscientific.com/worldscibooks/10.1142/10652 }

\begin{abstract}
Understanding the formation of earliest supermassive black holes is a question of prime astrophysical interest. In this chapter, we focus on the formation of massive black holes via gas dynamical processes. The necessary requirement for this mechanism are large inflow rates of about 0.1 solar mass per year. We discuss how to obtain such inflow rates via an isothermal collapse in the presence of atomic hydrogen cooling, and the outcome of such a collapse from three dimensional cosmological simulations in subsection 2.2.  Alternatives to an isothermal direct collapse are discussed in subsection 3 which include trace amounts of metals and/or molecular hydrogen. In the end, we briefly discuss future perspectives and potential detection of massive black hole seeds via upcoming missions. 
\end{abstract}


\body

\section{Introduction}
Understanding the formation of supermassive black holes remains an open and fascinating issue, as outlined in detailed reviews on this topic by \citet{Volonteri10,Volonteri2012,Haiman13,LatifFerrara2016}. The three main pathways for the formation of supermassive black holes are: (1) stellar remnants, (2) seed black holes forming  in dense stellar clusters via dynamical processes, (3) monolithic collapse of proto-galactic gas cloud into a massive black hole so-called direct collapse scenario. In case of stellar remnants, the most promising pathway is likely in the context of massive Population III stars, as introduced in the previous chapter 4. Such a mechanism, if it was to produce the observed supermassive black holes, would certainly require a mechanism of super-Eddington accretion, as described in chapter 11, which may however also be present in other scenarios. Black hole formation via collisions in stellar clusters will be discussed in chapter 7, and the observed masses and constraints on the supermassive black hole population at high redshift are reported in chapter 12. In this chapter, the main focus will be on black hole formation via a monolithic collapse. 

The idea for the formation of  a massive black hole directly via the gas dynamical processes was conceived in the pioneering work of Martin Rees \citep{Rees1984}.  The expectation is that  gas in the low spin halos collapses on the viscous time scale, forms a  rotationally supported compact disk  which later may lead to the formation of a massive black hole (BH) \citep{Loeb1994,Eisenstein1995}.  Similarly, models proposed in the cosmological context  suggest that conditions for the formation of a massive black hole are ideal  in high redshift protogalaxies with the lowest angular momentum gas \citep{Koushiappas2004, Volonteri2005,Lodato2006}.  It is proposed that  gas can rapidly loose angular momentum via 'bars within bars' instabilities  and may lead to rapid formation of a self-gravitating core supported by gas pressure. Such a core catastrophically cools by thermal neutrino emission and  contracts  to potentially form  a massive BH \citep{Begelman2006}. The first smoothed particle hydrodynamical simulations starting from idealised initial conditions showed that a low spin metal free halo  collapses into a single clump while higher spin halo formed a binary \citep{Bromm03}.

The key requirement for this scenario is that gas should rapidly collapse by efficiently transporting angular momentum and avoiding fragmentation. The goal is to bring large inflows ($\rm \geq 0.1~M_{\odot}/yr$, see discussion below) of gas into the halo centre within a short time scale of about $\sim$1 Myr and rapidly build up  a massive object of $\rm 10^4-10^6~M_{\odot}$.  Such large inflow rates can be obtained thermodynamically by keeping the gas warm as the inflow rate is $\propto  T^{3/2}$ and also via dynamical processes such as 'bars within bars' instabilities \citep{Shlosman1989, Begelman2006} or in the aftermath of galaxy mergers \citep{Mayer2010}. Depending on the time evolution of mass inflow rates, the central object may form a supermassive star/quasi-star (details are mentioned below) or directly collapse into a massive black hole.

\section{Black hole formation in primordial atomic gas}
The mass inflow rate ($\dot{M}$) of collapsing gas is related to its thermodynamical properties, as $\rm \dot{M} \sim c_s^3/G  \sim 0.1 ~M_{\odot}/yr \left( \frac{T}{8000 ~K}  \right)^{3/2} $ where $c_s$ is the sound speed and T is the gas temperature.  The higher the sound speed the larger the mass inflow rate. Therefore,  the thermodynamical requirement for getting large  inflows is that gas should not cool down to lower temperatures, otherwise it will fragment and form ordinary stars.  The cooling ability of the gas strongly depends on its chemical composition. In the presence of a trace amount of dust/metals, the gas  cools down to a few tens of Kelvin by radiating away its thermal energy and forming stars. Even in primordial gas, molecular hydrogen cooling  can bring the gas temperature down to $\rm \sim 200$ K and induces star formation.  In the absence of molecular hydrogen, primordial gas remains in the atomic phase, cools mainly via atomic line radiation and the gas temperature remains around 8000 K.  In atomic primordial gas collapse is expected to proceed isothermally with $\rm T \sim 8000~K$ and large mass inflow rates of the order of $0.1 ~M_{\odot}/yr$ can be achieved easily. Therefore, the conditions for forming a massive object are ideal in massive primordial halos with $T_{vir} \geq 10^4$ K cooled only via atomic lines. Such  halos have masses of a few times $\rm \geq 10^7~M_{\odot}$, formed at $z>10$ and their gravitational potentials are sufficiently deep to allow the rapid collapse. Therefore, massive primordial halos deprived of $\rm H_2$ cooling are the potential cradles for the formation of massive black holes.

The prerequisites for the formation of massive black holes in  atomic cooling halos are that they  should be metal-free and the formation of molecular hydrogen remains suppressed. In the next sub-section, we discuss in detail how to quench the molecular hydrogen formation which could be detrimental for forming massive black holes via isothermal direct collapse (DC). 

\subsection{Conditions to keep the gas primordial and atomic }
\begin{figure}
\hspace{-0.4cm}
\includegraphics[scale=0.65]{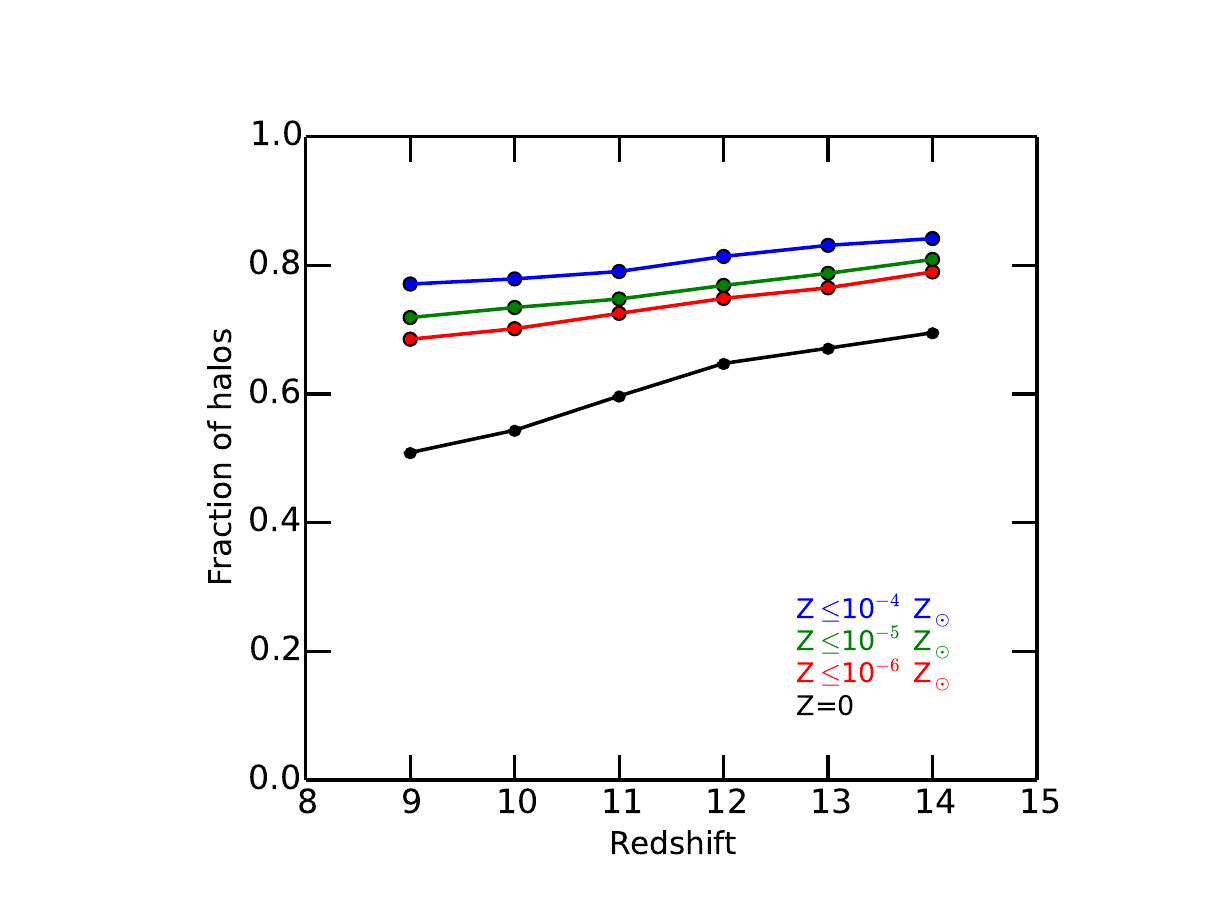}
\caption{ Fraction of halos with metallicity below the given value in the figure legged and masses between $\rm 2 \times ~10^7-10^8~M_{\odot} $.  Adopted from \citet{Latif2016D},  \textcopyright AAS. Reproduced with permission. }
\label{figure1}
\end{figure}

Our understanding of structures  in the cosmos is based on the hierarchical paradigm of structure formation according to which minihalos of $\rm 10^5-10^6~M_{\odot}$ were formed first at $z  \sim 30-40$ which later merged to form larger halos.  According to the Big Bang theory, the gas in the Universe initially has a primordial composition of predominantly hydrogen and helium. The first stars, so-called Population III stars, will then form out of a primordial gas in these minihalos.  As structure formation proceeds, depending on the their mass, these stars end their lives as supernovae and pollute the ambient medium with metals. At earlier cosmic times, a few hundred Myrs after the Big Bang, the metal enrichment is expected to be  patchy and also found  from the large scale numerical simulations \citep{Trenti2009,Maio2011,Ritter2014,Pallottini2014,Habouzit2016a}. Therefore, some of the halos may remain pristine until their masses reach the atomic cooling regime (above $10^7~M_{\odot}$). Indeed, there is observational evidence that pockets of metal free gas can exist down to z=7 \citep{Simcoe2012} and even at z=3 \citep{Fumagalli2011}.  The estimates about the fraction of metal free halos computed from cosmological simulations including self-consistently  star formation and supernova feedback suggest that about 40 \% of halos remain metal free down to z=10 \citep{Latif2016D, Habouzit2016b}, see figure \ref{figure1}.

As we discuss below, due to the requirement of a strong UV flux, a DC halo may form in the surroundings of an  intensely star forming galaxy. This may affect the abundance of DC halos by metal pollution from supernova winds  from a nearby star forming halo \citep{Dijksta2014,Habouzit2016a}. Even such pollution can be avoided in a synchronised pair of halos where the halo which is the radiation source forms first  while  a rapid collapse in the DC halo helps in avoiding the metal pollution \citep{Visbal2014}.  Moreover, cosmological hydrodynamical simulations starting from first principles show that metal ejection preferentially occurs in the low density regime \citep{Ritter2014,Pallottini2014} and neighboring halos  may remain metal free. In a nutshell,  metal pollution is not a bottleneck in the formation of DCBHs.

The second main constraint for DCBHs is that  the formation of $\rm H_2$ should remain suppressed in DC halos.  Trace amount of $\rm H_2$ can be formed via gas phase reactions where a residual fraction of electrons  from the recombination epoch acts as a catalyst. As discussed in the previous chapter on the chemistry of the early universe, the main pathway for the formation of $\rm H_2$ is the following:
\begin{equation}
\mathrm{H + e^{-} \rightarrow H^{-} +} \gamma
\end{equation}
\begin{equation}
\mathrm{ H + H^{-} \rightarrow  H_{2} + e^-.}\\ 
\label{h21}
\end{equation}

The $\rm H_2$ can be dissociated  either directly or indirectly by UV radiation depending on on the stellar spectra. The low energy photons with $\rm 0.75$ eV can photo-detach $\rm H^-$ which is the main channel for the formation of $\rm H_2$. The photons with energy between 11.2-13.6 eV, so-called Lyman Werner (LW) photons  directly photo-dissociate molecular hydrogen via the Solomon process. The dissociation processes are described by the following reactions:
\begin{equation}
\mathrm{H_{2}} + h\nu \mathrm{\rightarrow H + H}
\label{h20}
\end{equation}
\begin{equation}
\mathrm{H^{-}} + h\nu \mathrm{\rightarrow H + e^{-}}\\ 
\label{h2}
\end{equation}

\begin{figure}
\includegraphics[scale=0.65]{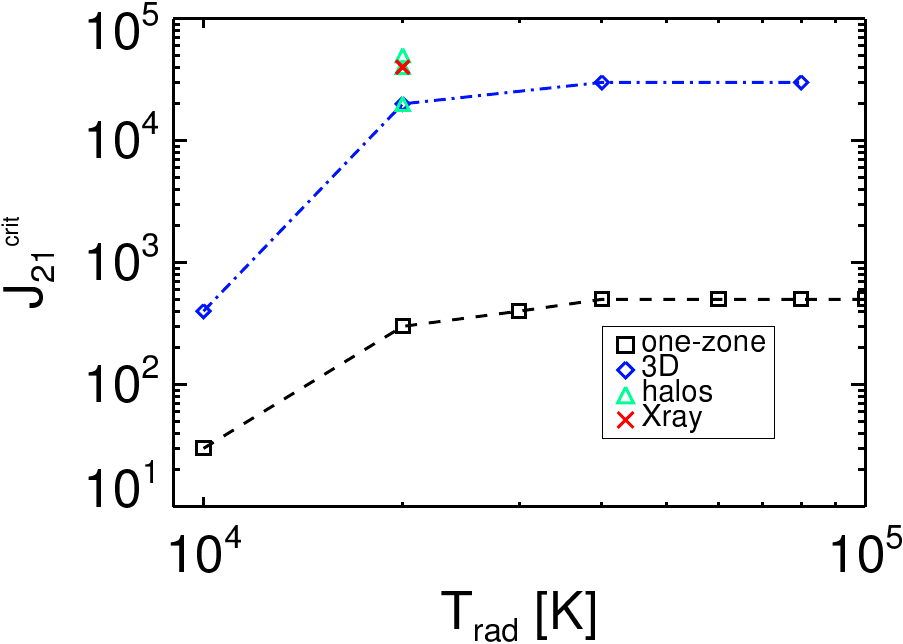}
\caption{Estimates of the critical value of the UV flux ($J_{21}^{\rm crit}$)  both from one zone models and 3D simulations including variations from halo to halo, dependence on  the radiation spectra and the impact of X-ray ionization. Adopted from  \citet{Latif2015}, reproduced by permission of Oxford University Press / on behalf of the RAS.  }
\label{figure2}
\end{figure}

The competition between the formation and dissociation timescales defines the  critical value of UV flux (hereafter $J_{21}^{crit}$) above which the formation of $\rm H_2$ remains quenched. Pop. III stars with $T_{rad}=10^5$ K produce more high energy photons and are very effective in directly dissociating $\rm H_2$ while normal stars with $\rm T_{rad}=10^4$ K photo-detach $\rm H^-$. The previous studies employed idealized spectra  to  compute the $J_{21}^{crit}$ and found that it varies from 30-300 for $\rm T_{rad}=10^4$ K and  1000 for $\rm T_{rad}=10^5$ K \citep{Omukai2001,Shang2010,Latif2014,Johnson2014}. However, recent estimates of  $J_{21}^{crit}$ for a realistic spectra of the first galaxies suggest that  it has no single value but it depends on the stellar age, metallicity and mode of star formation \citep{Sugimura2014,Agarwal2015,Agarwal2016}. Moreover, such spectra can be mimicked with $\rm T_{rad}=2 \times 10^4-10^5$ K. The values of $J_{21}^{crit}$  from cosmological simulations for a realistic spectra of first galaxies found that it further  depends on the properties of the host halos and varies between 20,000-50,000 considering a uniform background UV flux \citep{Latif2015} or anisotropic flux \citep{Regan2014B,Regan15b}. X-rays  catalyse $\rm H_2$ formation by boosting electron fraction and consequently the value of  $J_{21}^{crit}$ may further get enhanced \citep{Latif2015,Inayoshi2015,Regan2016}, see figure \ref{figure2}. The accurate treatment of $\rm H_2$ self-shielding is also necessary to calculate  $J_{21}^{crit}$ \citep{WolcottGreen2011,Hartwig2015}.

Estimate of $J_{21}^{crit}$ and the fraction of metal free halos are crucial to estimate the the number density of direct collapse black holes and to assess the feasibility of this scenario. A detailed assessment of its dependence on the stellar spectra reflecting our present understanding is therefore provided in chapter 6.
 
\subsection{Outcome of an isothermal collapse}
\begin{figure}
\hspace{0.8 cm}
\includegraphics[scale=0.5]{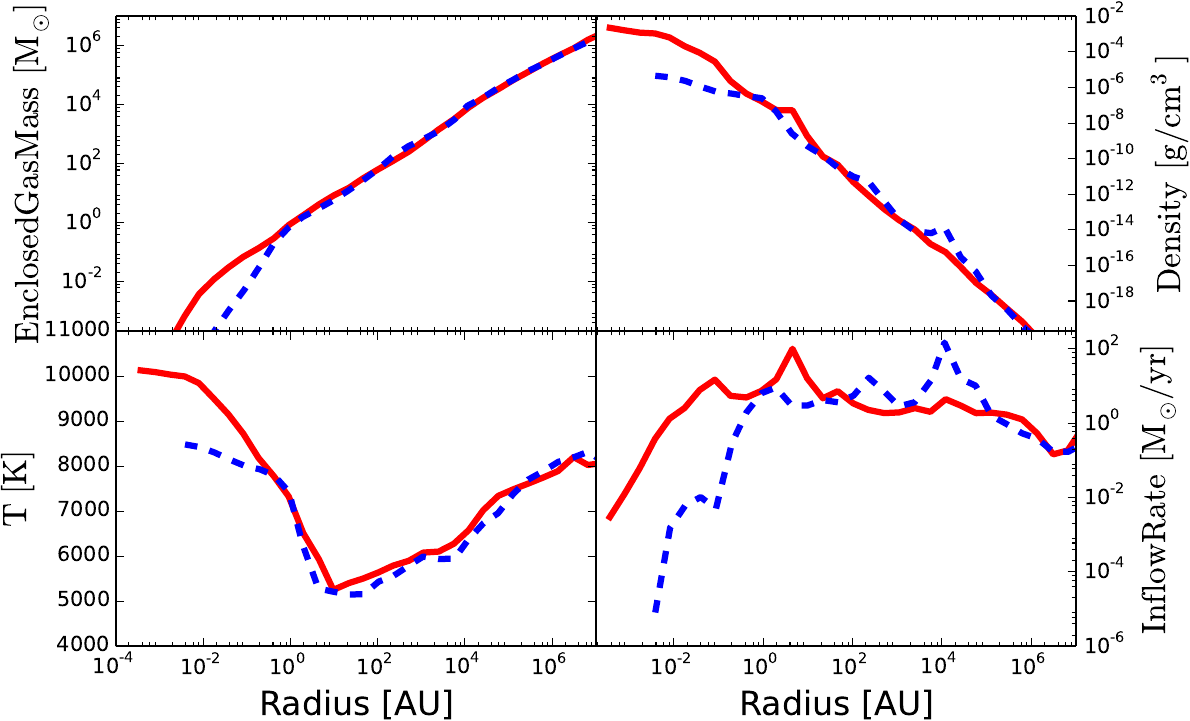}
\caption{Spherically averaged and radially binned profiles of density, temperature, mass and mass inflow rates are depicted here.  The simulated halos are metal free and illuminated by a strong LW flux above the $J_{21}^{crit}$. Consequently, $\rm H_2$ formation remains suppressed and cooling proceeds via atomic lines.  Adopted from \citet{Latif2016}, reproduced by permission of Oxford University Press / on behalf of the RAS. }
\label{fig3}
\end{figure}

In this section, we discuss the  outcome of an isothermal collapse  in massive primordial halos with $T_{vir} \geq 10^4$ K  illuminated by a strong LW flux above the critical strength.  Numerical cosmological simulations performed under these conditions confirm that the gas collapses isothermally in dark matter potentials with $\rm T \sim 8000$ K in a self-similar way and the density profile follows an $R^{-2}$ behaviour \citep{Bromm03,Wise2008,Regan09,Latif2011,Latif2013c,Inayoshi2014,Bcerra2014,Latif2016}.  The gas continues to cool  by Lyman alpha radiation  and keeps collapsing. At  densities of $\rm 10^{8}~cm^{-3}$, cooling due to $\rm H^-$ comes into play and brings  the gas temperature down to 5000 K \citep{VanBorm2014,Latif2016}. Above $\rm 10^{16}~cm^{-3}$,  the gas cloud becomes optically thick to both Lyman alpha and $\rm H^-$ cooling, and the temperature starts to rise as shown in figure \ref{fig3}.  The collisional ionization cooling becomes important and maintains the gas temperature close to $\rm 10^4$ K. Eventually, at densities higher than $\rm 10^{20}~cm^{-3}$, the gas cloud becomes completely opaque and collapses adiabatically.

Under isothermal conditions, the gas is expected to collapse monolithically without fragmentation. Previous  low resolution studies employing  a fixed Jeans resolution of four cells confirmed this hypothesis \citep{Regan09,Latif2011}. However, recent work employing a detailed chemical model,  resolving the collapse to unprecedentedly high densities of $\rm 10^{21}~cm^{-3}$  with a Jeans resolution of 32 cells shows that fragmentation occasionally occurs depending on the properties of host halos,  but does not prevent the formation of a massive central object \citep{Bcerra2014,Latif2016}.  The clumps forming due to the fragmentation on small scales quickly migrate inwards and  merge with the central object \citep{Inayoshi2014b,Latif2015Disk}. Analytical models of  primordial disks around the central object suggest that  in the presence of large inflows and rapid rotation, viscous heating stabilises the disk and helps in the formation of a massive central object \citep{Latif2015Disk2,Schleicher2016}. Similarly,  magnetic fields amplified via the so-called small scale dynamo reach the equipartition field strength within  a dynamical timescale and provide a support against gravity. Such strong fields further help in suppressing the fragmentation in atomically cooled halos \citep{Schobera,Schliecher2010dyn, Latif2013a,LatifMag2014}. On larger scales, the angular momentum in these halos gets transferred due to the gravitational torques exerted by the triaxility of DM haloes via 'bars within bars'  instabilities \citep{Choi2013,Choi2015}.

\begin{figure}
\hspace{-0.0 cm}
\includegraphics[scale=0.2]{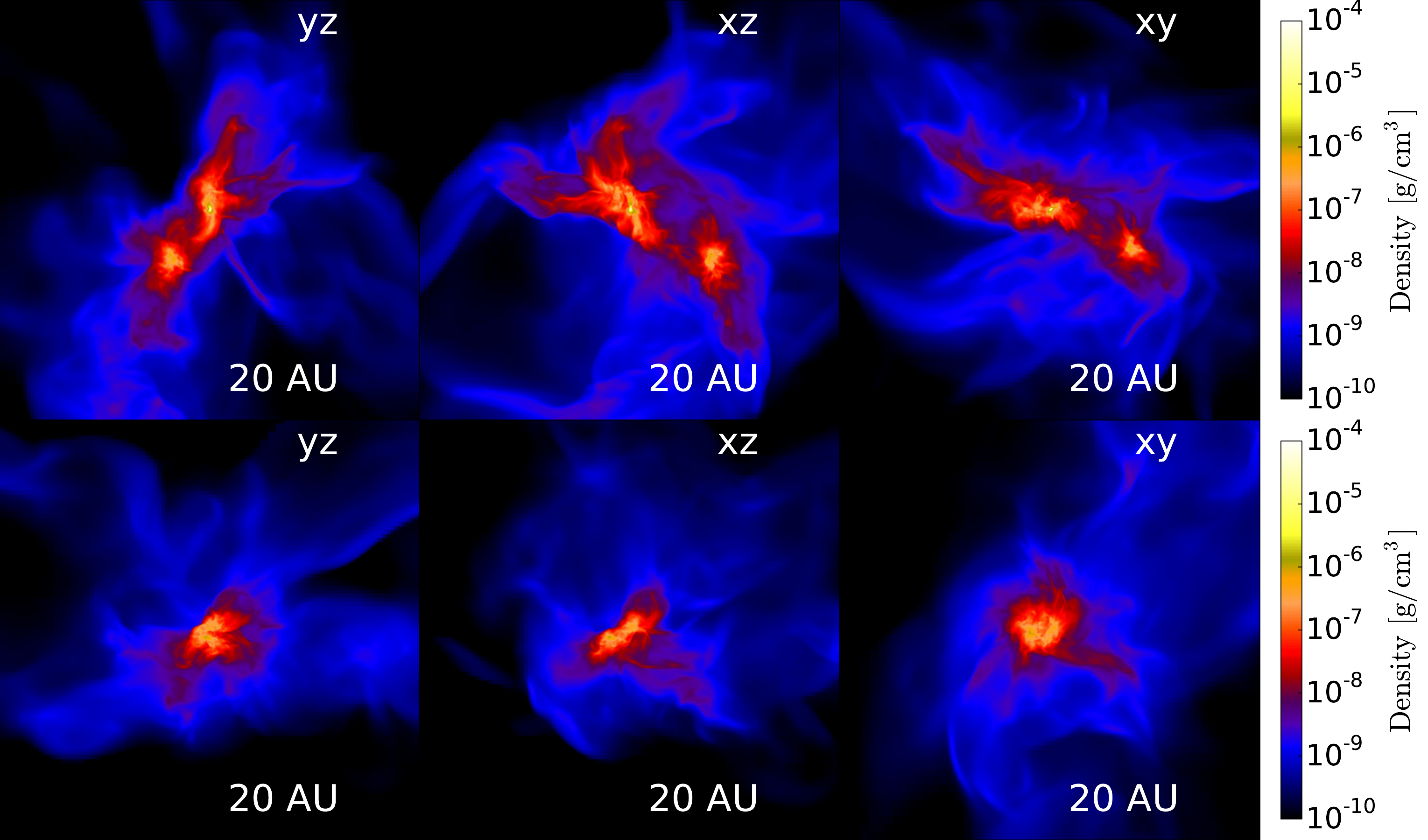}
\caption{Averaged density along the line of sight in the central 20 AU for two halos (top and bottom). The simulated halos are metal free and illuminated by a strong LW flux above the $J_{21}^{crit}$. Therefore, cooling proceeds via atomic lines and an isothermal collapse occurs.  Adopted from \citet{Latif2016}, reproduced by permission of Oxford University Press / on behalf of the RAS. }
\label{fig4}
\end{figure}

One of the salient features  of forming DCBHs through isothermal collapse is that large inflow rates of $\rm 0.1-1~M_{\odot}/yr$ are easily available, see bottom right panel of figure \ref{fig3}. If these inflow rates can be sustained for about one Myr then the formation of a massive object of  $\rm \sim 10^4 - 10^5 ~M_{\odot}$ is feasible. In fact, it has been confirmed from cosmological numerical simulations that a massive central object of $\rm 10^5~M_{\odot}$ can be formed within a Myr after the initial collapse \citep{Latif2013d,Shlosman2016}, though these simulations did not employ the feedback from a supermassive proto-star.  In fact, stellar evolution calculations show that for large mass accretion rates $\dot{m}$)  of $\geq 0.1~M_{\odot}/yr$, the radius of the star monotonically increases with mass because of the short accretion time compared to the Kelvin Helmholtz contraction timescale \citep{Hosokawa12,Schleicher13}. These stars have low surface temperatures of about 5000 K like supergaints and  do not produce strong UV flux. Due to the short accretion  time in comparison with the nuclear burning time, the core of such a supermassive star may collapse into a BH provided that sufficiently large accretion rates can be retained \citep{Begelman2008,Begelman2010}. Such stars with a BH at their centre are called quasi-stars and it has been found that $\dot{m} >0.14~M_{\odot}/yr$ is required for their formation \citep{Schleicher13}. These supermassive/quasi-stars may collapse into a BH via general relativistic instabilities \citep{Volonteri2010b,Ball2011,Johnson2014,Ferrara14} and are the potential embryos of massive black holes forming via direct collapse. A more detailed picture on the evolution of such supermassive stars is provided in chapter 8.

Most of the previous work studying the  evolution of a supermassive star used constant mass accretion rates. However, recent studies employing time dependent accretion rates show that if the time interval between to consecutive episodes of accretion ($\Delta t_{acc}$) is longer than 1000 years  then star can sufficiently contract and produces strong UV flux which may halt further accretion  \citep{Sakurai2015}.  It is expected that accretion onto the supermassive protostar will be episodic due to  the possible fragmentation of proto-stellar  disk and clumps inward migration. Under isothermal conditions, $\Delta t_{acc}$ is expected to be  much shorter in 1000 years  \citep{Sakurai2016,Latif2015Disk2,Latif2016}. Therefore, intermittent accretion does not halt the formation of a potential supermassive star.  As mentioned earlier, due to the requirement of a strong LW flux, the host halos of DCBHs are expected to form in the close vicinity of actively star forming galaxy and  some of these DC halos get tidally disrupted by the nearby massive halos and can not host  a DCBH \citep{chon16}. Recently, radiation hydrodynamical simulations investigating the impact of ionising radiation on the formation of DCBHs found that rich structures such as filaments and clumps between DC halo and star forming galaxy shield it from ionising photons \citep{Chon2017}. So, the formation of DCBHs is expected to continue under these conditions.

\subsection{Outcome of  collapse  under less idealized conditions}
Various alternatives to isothermal direct collapse mediated by a strong LW flux have been proposed.  For  large columns ($\rm \geq 10^{20}~cm^{-2}$) of neutral hydrogen, the escape time for Lyman alpha photons becomes longer  than the cloud collapse time scale. Consequently, Lyman alpha photons  get trapped inside the  cloud, the equation of state gets stiffened, the temperature of the gas cloud starts to increase and eventually collapse proceeds adiabatically \citep{Spaans2006}. However, later studies found that cooling still proceeds via 2s-1s transition and collapse becomes  isothermal  identical to the Lyman alpha cooling case \citep{Schleicher10,Latif2011b}. Similarly, large baryonic streaming motions (3 $\sigma$ fluctuations) produced prior to the epoch of recombination  naturally increase the critical mass for $\rm H_2$ cooling until it reaches the atomic cooling limit. Such motions may also collisionaly dissociate $\rm H_2$ molecules and avoid metal enrichment by suppressing  in-situ star formation \citep{Tanaka2013,Tanaka2014}. However, it was found that streaming motions are not effective in quenching $\rm H_2$ formation and may require the ubiquity of a strong LW flux \citep{Latif2014Stream}.

The large inflow rates of the gas required to assemble massive back holes can be obtained dynamically  via 'bars within bars' instabilities,  and fragmentation even in metal rich gas may be suppressed via supersonic turbulence \citep{Begelman2009}. However, studies of contemporary star formation show that supersonic turbulence locally compresses the gas and induces star formation in molecular clouds \citep{Federrath2010,Federrath11}. During galaxy mergers, gravitational torques drive large inflows of about $10^3-10^4~M_{\odot}/yr$ to the halo centre and form a compact circumnuclear disk which later may coalesce into a massive BH \citep{Mayer2010}. The cooling timescale of such disk is very short, of the order of 100 years, and hence the resulting mass scale is still controversial \citep{Ferrara13}. In a recent study, improving on their previous work of galaxy mergers, \citet{Mayer2015} found that the gas becomes optically thick to cooling radiation and the central stable core may directly collapse  into a massive BH via general relativistic instabilities. In this scenario, the central core collapse must avoid the stellar phase otherwise it will blow away  most of the  mass via stellar winds as the mass loss from star is directly proportional to the metallicity of  the gas.

So far, we have mainly focused on the isothermal direct collapse with emphasis on keeping  the halo metal free and  devoid of $\rm H_2$.  The studies of primordial star formation suggest that  although the protostellar disk forming in natal primordial clouds fragments, most of the clumps migrate inward, leading to intermittent accretion and  merging with the central protostar. The typical accretion rates in primordial minihalos are $\rm 10^{-2}-10^{-4}~M_{\odot}/yr$ which is about 3-4 orders of magnitudes larger than the present day star formation in molecular clouds. Moreover,  the accretion rate in the bursty mode exceeds $\rm 10^{-2}~M_{\odot}/yr$, which keeps the stellar envelope bloated up and consequently the supermassive star produces weak UV feedback.  Thus, even in the presence of $\rm H_2$,  massive stars up to $\rm 1000 ~M_{\odot}$ or even higher may form \citep{Hirano2014,Latif2015Disk,Hosokawa2016}. 

\begin{figure}
\hspace{0.8 cm}
\includegraphics[scale=0.6]{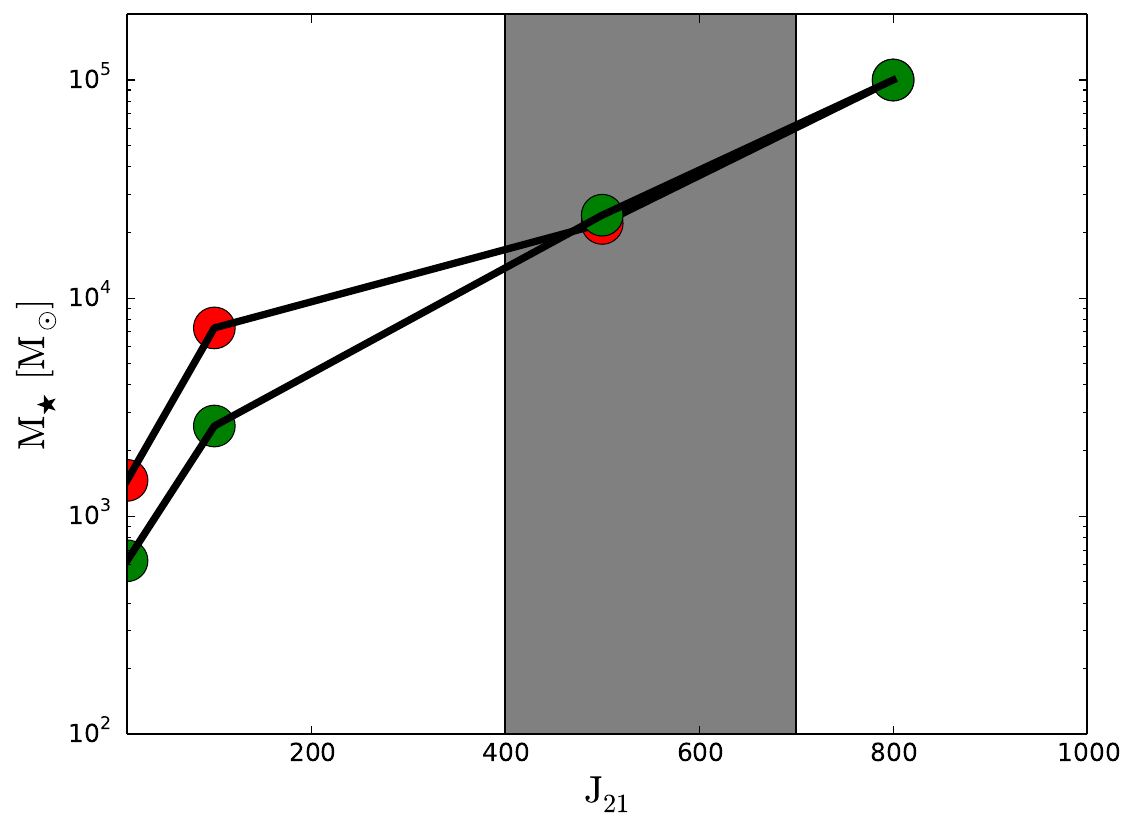}
\caption{Expected stellar masses for different  strengths of  background LW flux. The dashed vertical shaded region represents the range of the critical values. The blue and red spheres two representative halos. Adopted from \citet{Latif2014ApJ},  \textcopyright AAS. Reproduced with permission. }
\label{fig5}
\end{figure}

The numerical experiments exploring  collapse in massive primordial halos with $\rm T_{vir} \geq 10^4$ K irradiated by  moderate strength of LW flux found  that a trace amount of molecular hydrogen forms in the halo centre surrounded by warm gas with $T \sim 8000$ K. It was found that fragmentation occasionally occurs but clumps migrate inward and merge with the central star.  Large inflow rates of $\rm \sim 0.1~M_{\odot}/yr$ are generated in halos which have recently gone through a major  merger \citep{Latif2014ApJ,LatifVolonteri15}. Under these conditions, the central star may reach $\rm 10^4~M_{\odot}$ within 1 Myr even if  in situ star formation occurs in these halos.  Estimates about the mass of the central object for various strengths of LW flux below $J_{21}^{crit}$ are shown in figure \ref{fig5}. These findings suggest that a massive central object/star of $\rm 10^3-10^4~M_{\odot}$  can be formed within 1 Myr after the initial collapse for moderate LW flux. Even if in situ star formation occurs,  the gas flow to the central star may still continue until stars reach the main-sequence and below away the gas. This study was performed assuming that the spectra of Pop. II stars is soft with $\rm T_{rad}= 10^4~K$. However, similar results are expected for a realistic spectra of Pop. II stars.

\begin{figure}
\hspace{0.8 cm}
\includegraphics[scale=0.2]{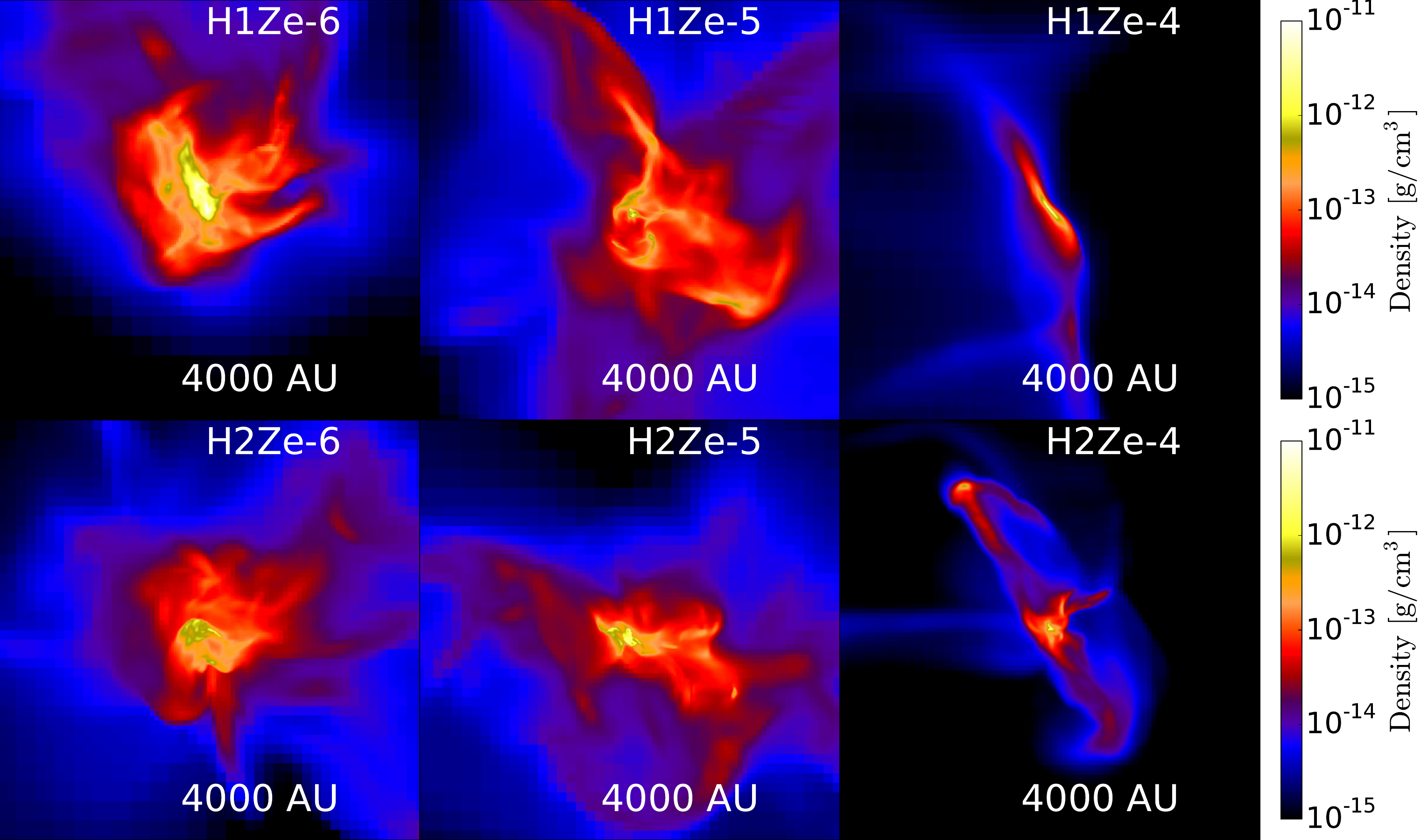}
\caption{ Average gas density along x-axis for metallicities of $\rm Z/Z_{\odot}=10^{-6}, 10^{-5}, 10^{-4}$ and is shown for the central 4000 AU of a halo. Each row represents a halo (halo 1 on top and halo 2 on bottom) and each column represents metallicity (increasing from left to right). Adopted from \citet{Latif2016D},  \textcopyright AAS. Reproduced with permission.}
\label{fig6}
\end{figure}

In the presence of a strong LW flux ,star formation within the halo remains suppressed but the halo may be polluted by supernova winds from a nearby star forming galaxy.  It has been found that  for  a trace amount of metals/dust as low as $Z = 10^{-6}Z_{\odot}$, where Z is the metallicity of the halo and $Z_{\odot}$ is the solar metallicity, dust  cooling becomes important at densities above $\rm 10^8-10^{12}~cm^{-3}$  and lowers the gas temperature to a few hundred K \citep{Omukai2008,Latif2016D}. For very low metallicities $\rm \leq 10^{-5}$ and a strong LW flux, cooling remains confined to the central 10 AU and large inflows of gas are available.  The density structure for two halos  illuminated by a strong LW flux and polluted by a trace amount of metals are shown in figure \ref{fig6}.  It has been found that  for $\rm Z/Z_{\odot}= 10^{-4}$, due the efficient dust cooling, a filamentary structure emerges and gravitationally bound clumps form. While for $\rm Z/Z_{\odot}= 10^{-5}$, the density structure remains spherical and gravitationally unbound sub-solar clumps are observed.  The clumps for the cases with $\rm Z/Z_{\odot} \leq 10^{-5}$ cases may migrate inwards and merge with the central object. In the case of efficient fragmentation, a dense stellar cluster is expected to form as fragmentation occurs in the very inner part of the halo. It is expected that the run-away collisions in such a dense cluster may lead to the formation of a very massive central star which may later collapse into a massive BH of about a thousand solar masses. So, the conditions for growing a massive object in very metal poor halos with  $\rm Z  \leq 10^{-5}~Z_{\odot}$ and illuminated by a strong UV flux are still favourable \citep{Latif2016D}.

\section{Future outlook}
Despite the tremendous progress made regarding our understanding of massive black hole formation during the past decade, there are still many open questions. It is yet not clear what is  the final outcome of an isothermal collapse, if it is always a supermassive star or  a quasi-star or a direct massive black hole.  The final result depends  on the long term sustainability of the mass accretion rates and the amount of angular momentum retained during the end stages of gravitational collapse.  Both of these quantities remain uncertain and are not fully comprehended due to the numerical constraints. In the future, numerical simulations starting from ab initio initial conditions should be performed to assess the feasibility of both parameters and to determine the ultimate fate of direct collapse.  Similarly,  stellar evolution calculations of supermassive proto-star formation ignored the role of rotation which may impede the collapse and may shut further mass accretion.

Some studies suggest that rapid accretion may launch strong winds leading to large mass loss \citep{Dotan2011,Fiacconi2016}. Also, preliminary work exploring the role of rotation during the quasi-stellar phase indicates that  the quasi-stellar phase may be skipped for objects massive than $\rm 10^5~M_{\odot}$ \citep{Fiacconi2017}. However, detailed three-dimensional radiation hydrodynamical computations are necessary to support/reject this finding.  The dynamical ways to obtain large inflow rates look promising, but it is not completely clear whether such inflows can be maintained down to AU scales.  Also, whether metal-enriched gas remains optically thick or radiates away thermal energy to form stars remains to be determined. Detailed investigations exploring this channel will thus be necessary.

 One of the biggest uncertainty in assessing the feasibility of direct collapse black holes is their number density which sensitively depends on the value of $\rm J_{21}^{crit}$.  The variation in $J_{21}^{crit}$ by a factor of a few changes the abundance of direct collapse black holes by an order of magnitude or even larger. Our current understanding suggests that  values of $J_{21}^{crit}$ vary by orders of magnitude and may also change with the age of stars and metallicity. In the future, more realistic estimates of $\rm J_{21}^{crit}$ are required  from 3D radiation hydrodynamical simulations by properly modelling the SED of the source galaxy for different modes of star formation, as also discussed in chapter 6.

 Observational evidence is  necessary to  constrain these models of  black hole formation. The first observations of CR7, the brightest Lyman alpha emitter at z=6.6, revealed that it shows strong Lyman alpha and He-1640$~\AA$ emission and no signatures of  metal lines were observed \citep{Sobral}. The drivers of  strong Lyman alpha and He-1640$\AA$  can be  either a cluster of primordial stars with  $\rm 10^7~M_{\odot}$ or a direct collapse black hole of $\rm 10^6-10^7~M_{\odot}$.  However, due to the metal enrichment  forming such a young massive cluster of primordial stars at redshifts as low as z= 6 seems infeasible and a  black hole of $\rm 10^6~M_{\odot}$ possibly forming via DC is more likely source of CR7 \citep{Pallottini2015,Hartwig16,Agarwal16,Dijkstra2016,Smidt2016,Smith2016}.  The recent deep observations of CR7 show doubly ionized oxygen (OIII) emission \citep{Bowler17} but the current photometry is still  consistent with a mild amount of pollution  DCBH site with metals \citep{Hartwig16,Pacucci2017,Agarwal2017}.  It is expected that JWST\footnote{https://jwst.nasa.gov} will be able to directly probe the observational signatures of DCBHs in high redshift galaxies.
 
 The future X-ray space observatory ATHENA\footnote{http://www.the-athena-x-ray-observatory.eu} is expected to detect a few hundred low luminosity active galactic nuclei (AGN) in the early universe with X-ray luminosities of $\rm L_{X} \geq 10^{43} erg/s$. These observations will provide direct constraints on the masses and luminosity functions of  the first AGN.  Also future 21 cm experiments such as SKA\footnote{http://skatelescope.org} and LOFAR\footnote{http://www.lofar.org/} will probe the proximity zones of high redshift quasars and help in better understanding their formation and growth mechanisms \citep{Whalen2017}. In the local universe the deep observation of dwarf galaxies may help in tracing the seed BHs forming at high redshift \citep{Reines2016}. The recent detections of BH-BH mergers with LIGO\footnote{https://www.ligo.caltech.edu} have opened a new window of gravitational waves astronomy and the upcoming European space mission eLISA\footnote{https://www.elisascience.org} will be able to detect the merging of massive black holes and help in understanding their formation mechanisms.

In the next chapters, we will clarify uncertainties in the value of $\rm J_{21}^{crit}$ (chapter 6), which is crucial to quantify the expected number of black holes. Black hole formation through mergers in stellar clusters will then be explored in chapter 7, and the evolution of supermassive stars is described in chapter 8. The growth of seed black holes, potentially forming from the first stars, will be discussed in chapters 10 and 11. Statistical predictions will be provided in chapter 9. A comparison with current observations of high-redshift quasars is given in chapter 12, while predictions on gravitational wave emission are provided in chapter 13, and expectations for future observations are outlined in chapter 14.

{
\bibliographystyle{ws-rv-har}    
\bibliography{ref}
}

\printindex[aindx]           
\printindex                  

\end{document}